\documentstyle[12pt,fleqn]{article}
\def\la{\mathrel{\mathpalette\fun <}}
\def\ga{\mathrel{\mathpalette\fun >}}
\def\fun#1#2{\lower3.6pt\vbox{\baselineskip0pt\lineskip.9pt
        \ialign{$\mathsurround=0pt#1\hfill##\hfil$\crcr#2\crcr\sim\crcr}}}
\begin{document}
\begin{titlepage}
\null\vspace{-72pt}
\begin{flushright}
{\footnotesize
FERMILAB--Pub--96/014-A\\
astro-ph/9601096\\
January 1996 \\
Submitted to {\em Phys.\ Rev.\ D}}
\end{flushright}
\renewcommand{\thefootnote}{\fnsymbol{footnote}}
\vspace{0.15in}
\baselineskip=24pt

\begin{center}
{\Large \bf  Eternal annihilations of light photinos}\\
\baselineskip=14pt
\vspace{0.75cm}
Edward W.\ Kolb\footnote{Electronic mail: {\tt rocky@rigoletto.fnal.gov}}\\
{\em NASA/Fermilab Astrophysics Center\\
Fermi National Accelerator Laboratory, Batavia, IL~~60510, and\\
Department of Astronomy and Astrophysics, Enrico Fermi Institute\\
The University of Chicago, Chicago, IL~~ 60637}\\
\vspace{0.4cm}
Antonio Riotto\footnote{Electronic mail:
{\tt riotto@fnas01.fnal.gov}}\\
{\em NASA/Fermilab Astrophysics Center\\
Fermi National Accelerator Laboratory, Batavia, IL~~60510}\\
\vspace{0.4cm}
\end{center}

\baselineskip=24pt

\begin{quote}
\hspace*{2em} In a class of low-energy supersymmetry models the
photino is a natural dark matter candidate.  We investigate the
effects of post-freeze-out photino annihilations which can generate
electromognetic cascades and lead to photo-destruction of $^4$He and
subsequent overproduction of D and $^3$He. We also generalize our
analysis to a generic dark matter component whose relic abundance is
{\it not} determined by the cross section of the self-annihilations
giving rise to electromagnetic showers.
\vspace*{8pt}

PACS number(s): 98.80.Cq, 98.80.Ft, 14.80.Ly

\renewcommand{\thefootnote}{\arabic{footnote}}
\addtocounter{footnote}{-2}
\end{quote}
\end{titlepage}

\newpage

\baselineskip=24pt
\renewcommand{\baselinestretch}{1.5}
\footnotesep=14pt

\vspace{36pt}
\centerline{\bf I. INTRODUCTION}
\vspace{24pt}

\def\beq{\begin{equation}}
\def\eeq{\end{equation}}
\def\beqa{\begin{eqnarray}}
\def\eeqa{\end{eqnarray}}
\def\tr{{\rm tr}}
\def\ph{\tilde{\gamma}}
\def\g{\tilde{g}}
\def\x{{\bf x}}
\def\p{{\bf p}}
\def\k{{\bf k}}
\def\z{{\bf z}}
\def\re#1{{[\ref{#1}]}}
\def\eqr#1{{Eq.\ (\ref{#1})}}

There are good reasons to consider models of low energy supersymmetry
\re{haber} in which dimension-3 supersymmetric breaking operators are
highly suppressed. The low-energy features and possible new signatures
of such attractive form of supersymmetric breaking have been
extensively outlined by Farrar \re{glennys}. In most of these models
photinos $\ph$ and gluinos $\g$ are very light and the lightest
$R$-odd particle\footnote{$R$-parity is a multiplicative quantum
number, exactly conserved in most of supersymmetric models, under
which ordinary particles have $R=1$ while their superpartners have
$R=-1$.} may be a color-singlet state containing a gluino, the $R^0$,
with a mass $m_{R^0}$ in the $1$ to $2$ GeV range.

It has been recently pointed out by Farrar and Kolb \re{kolb} that a
photino $\ph$ slightly lighter than the $R^0$, in the mass range 100
to 1400 MeV, would survive as the relic $R$-odd species and might be
an attractive dark-matter candidate. Indeed, they found that it is
crucial to include the interactions of the photino with the $R^0$. The
$R^0$ has strong interactions and thus annihilates extremely
efficiently and remains in equilibrium to temperatures much lower than
its mass. In these circumstances, photino freeze-out is {\it no
longer} determined by self-annihilations into fermions pairs
$\ph\ph\longleftrightarrow f\bar{f}$ (which would result in an
unacceptable photino relic abundance for such low mass range), but
occurs when the rate of reactions converting photinos to $R^0$'s falls
below the expansion rate of the Universe. The rate of the $\ph-R^0$
interconversion interactions which keep photinos into equilibrium
($\ph\pi\longleftrightarrow R^0\pi$) or $R^0$ decay/inverse decay
($\ph \pi\longleftrightarrow R^0$), depends upon the densities of
photinos and pions, rather then on the square of the photino density,
as in the case for the self-annihilation processes. For photinos in
the relevant mass range ($m_{\ph}\sim 800$ MeV), the pion abundance is
enormous compared to the photino abundance. Therefore the photinos
stay in equilibrium to much higher values of $x\equiv m_{\ph}/T$ than
they would if self-annihilation were the only operative process,
resulting in a smaller relic density for a given photino mass and
cross section. Light photinos with a mass in the range $1.2\la r\equiv
m_{R^0}/m_{\ph} \la 2.2$ are cosmologically acceptable and in the
range $1.6\la r \la 2$ are excellent dark matter candidate.

The aim of the present paper is to investigate whether in the scenario
outlined in \re{kolb}, light photino self-annihilations into fermions
pairs may have an impact on the successful predictions of the
standard big-bang nucleosynthesis even though self-annihilations are
not relevant in determining the relic photino abundance. The
destructive high-energy photons coming from the self-annihilation
products of primordial photinos, including $e^{\pm}$, $\mu^{\pm}$ and
$\gamma$'s, may indeed cause photofission of the primordially produced
light nuclei \re{lindley} upsetting the agreement between big-bang
nucleosynthesis and the observed element abundances.
 
According to the standard lore, the source of destructive high-energy
photons may be neglected because self-annihilations (SA) typically
freeze out at a temperature $T_{{\rm SA}}\sim m_{\ph}/15$ (not to be
confused with the freeze out temperature of the photino relic
abundance $T_*\sim m_{\ph}/(20-25)$ determined by $\ph-R^0$
conversion). For photino masses in the dangerous range, $m_{\ph}$
greater than a few MeV, this freeze-out occurs well before the light
elements first become vulnerable to photodissociation at temperatures
of about a few keV. By this late epoch, any photons and electrons
generated before freeze-out have harmlessly thermalized. However,
after freeze-out, occasional self-annihilations still may take
place. Although rare on the expansion time scale, residual
annihilations continue to produce high-energy photons well after
nucleosynthesis ends, thereby placing the survival of the light nuclei
in jeopardy \re{residual}.

We explore below the consequences of these eternal light photino
self-annihilations by computing the residual annihilation rate and the
corresponding photofission yields. The typical feature of the model at
hand is that the production rate of dangerous high-energy photons and
the relic abundance of light photinos are {\it not} determined by the
same cross section, the former being dependent on the light photino
thermally averaged self-annihilation cross section $\langle
\sigma_{{\rm SA}} |v|\rangle$ and the latter by the $\ph-R^0$
conversion rate.  We then generalize our analysis to any case in which
the relic abundance of a generic dark matter candidate $X$ is not
determined by its self-annihilation cross section, as assumed in
\re{residual}, but rather by some other generic processes, a typical
case being the presence of a slightly heavier particle $X^\prime$ ({\it e.g.,} the
$R^0$ in \re{kolb}) with relative interconversion processes
$X\longleftrightarrow X^\prime$, usually called co-annihilation
\re{griest}.

\vspace{36pt}
\centerline{\bf II. ETERNAL ANNIHILATIONS}
\vspace{24pt}

In the Boltzmann equation \re{kt} for the evolution of the
$\ph$-number density there are several terms, including photino
self-annihilation ($\ph\ph\rightarrow X$), co-annihilation ($\ph
R^0\rightarrow X$), inverse decay ($\ph \pi\rightarrow R^0$) and
photino-$R^0$ conversion $(\ph\pi\rightarrow R^0\pi$) (see Refs.\
[\ref{kolb},\ref{preparation}] for a detailed analysis).

In this Section we are only interested in the terms of the Boltzmann
equation which may provide a source for destructive electromagnetic
cascades at low temperatures, {\it i.e.,} in the terms accounting for
photino self-annihilation into fermion pairs and photino pair
production from light particles in the plasma.

Assuming the light annihilation products are in thermal equilibrium,
these terms are of the form
\begin{equation}
\label{bol}
\dot{n}_{\ph}+3 H n_{\ph} =-\langle \sigma_{{\rm SA}} |v|\rangle \left[
\left(n_{\ph}\right)^2-\left(n_{\ph}^{{\rm EQ}}\right)^2\right],
\end{equation}
where $H$ is the expansion-rate of the Universe
 \begin{equation}
H=1.66\:g_*^{1/2}\frac{T^2}{M_{{\rm Pl}}}=
4.4\times 10^{-19}\:x^{-2}\:\left(\frac{m_{\ph}}{{\rm GeV}}\right)
\:\:{\rm GeV},
\end{equation}
$g_*$ being the relativistic degrees of freedom at the temperature
$T$, $n_{\ph}^{{\rm EQ}}$ is the photino equilibrium number density at
temperature $T$
\begin{equation}
n_{\ph}^{{\rm EQ}}(T)
=\frac{2}{(2\pi)^{3/2}}\:x^{-3/2}\: m_{\ph}^3\:{\rm exp}(-x),
\end{equation}
and, finally, an overdot denotes a time derivative. 

Well before photinos freeze-out, at times $t\ll t_{{\rm SA}},t_*$
(dictated respectively by self-annihilation processes and $\ph-R^0$
conversion), the self-annihilation rate $\Gamma_{{\rm SA}}=n_{\ph}
\langle \sigma_{{\rm SA}} |v|\rangle$ is faster than the expansion
rate, $\Gamma_{{\rm SA}}\gg H$, and the photino abundance is kept at
its equilibrium value, $n_{\ph}=n_{\ph}^{{\rm EQ}}$. We are interested
in the epoch long after freeze-out, $t\gg t_*$, when $n_{\ph}\gg
n_{\ph}^{{\rm EQ}}$ and the number density to the entropy density
$Y_{\ph}=n_{\ph}/s$, $s=(2\pi^2/45)g_{*S}T^3$ being the entropy
density at the temperature $T$, has ceased to decrease significantly,
{\it i.e.}, $Y_{\ph}\simeq Y_\infty$ for $t\gg t_*$.  The asymptotic
value $Y_\infty$, which is determined by the photino interactions with
$R^0$ and does not depend on the details of $\langle\sigma_{{\rm
SA}}|v|\rangle$ \re{kolb}, may be expressed in terms of the fraction
of the critical density contributed by the photino at the present
epoch
\begin{equation}
\Omega_{\ph}h^2=\frac{\rho_{\ph}}{\rho_C}=2.8\times 10^8\:\left(\frac{m_{\ph}}{{\rm GeV}}\right)\:Y_\infty,
\end{equation}
where $\rho_C=1.05\times 10^{-5}\: h^2$ GeV cm$^{-3}$ is the present
critical density.

Well after the freeze-out of the photino abundance, the change in time
of the fraction of photino number density contributing to fermion pair
production is then given by\footnote{It is easy to check that for the
temperature of interest here, $T\ll T_*$, self-annihilations into
fermion pairs is the dominant source for photino number density
depletion so that \eqr{start} actually describes the change in time of
the whole photino number density.}
\begin{equation}
\label{start}
\dot{n}_{\ph}+3 H n_{\ph} =-\langle \sigma_{{\rm SA}} |v|\rangle
Y_{\infty}^2\:s^2.
\end{equation}
This last Equation will be our starting point in the next Sections to
analyze the influence of eternal self-annihilations on primordial
nucleosynthesis.

\vspace{36pt}
\centerline{\bf III. ELECTROMAGNETIC CASCADES}
\vspace{24pt}

Let us begin by examining the manner in which annihilations of light
photinos into light leptons (electrons or muons) generate
electromagnetic cascades in the radiation-dominated thermal plasma of
the early Universe \re{sarkar}.

The epoch of interest for cascade nucleosynthesis has temperatures
smaller than a few keV, by which time production of the light elements
D, He, and Li by big-bang nucleosynthesis has completed. At this
epoch, electron-positron pairs are no longer in equilibrium and
blackbody photons, $\gamma_{bb}$, constitute the densest target for
electromagnetic cascade development.

A cascade is initiated by a high-energy lepton (or photon) coming from
the self-annihilating photinos and develops rapidly in the radiation
field mainly by photon-photon pair production and inverse Compton
scattering
\begin{equation}
e+\gamma_{bb}\rightarrow e^\prime+\gamma^\prime,\:\:\:\:
 \gamma+\gamma_{bb}\rightarrow
e^{+} + e^{-}.
\end{equation}
When cascade photons reach energies too low for pair production on the
blackbody photons, the cascade development is slowed and further
development occurs in the gas by ordinary pair production, but with
electrons still loosing energy mainly by inverse Compton scattering in
the blackbody radiation
\begin{equation}
\gamma+Z\rightarrow Z+e^{+}+e^{-},\:\:\:\:e+\gamma_{bb}\rightarrow
 e^\prime+\gamma^\prime.
\end{equation}
At high energies, the cascade develops entirely on the blackbody
photons by photon-photon pair production and inverse Compton
scattering. The characteristic interaction rates for these processes
are much higher than the expansion rate and thus one can assume the
cascade spectrum is formed instantly. This spectrum is referred to as
the ``zero-generation'' spectrum.

Since $\gamma-\gamma$ elastic scattering is the dominant process in a
radiation-dominated plasma for photons just below the pair-production
threshold \re{elastic}, a primary photon or lepton triggers a cascade
which develops until the photon energies have fallen below the maximum
energy $E_{{\rm max}}(T)\simeq m_{{\rm e}}^2/(22\:T)$ corresponding to
the energy for which the mean free paths against $\gamma-\gamma$
scattering and $\gamma-\gamma$ pair production are equal \re{ellis}.

When $E_{{\rm max}}$ is less than the threshold for
photodisintegration of $^4$He nuclei (approximately 20 MeV), cascade
nucleosynthesis is inefficient. This condition restricts the epoch of
cascade nucleosynthesis to $T\la 0.5$ keV. At temperatures between about
$10^{-4}$ and 0.5 keV, where the subsequent cascade development
is via ordinary pair production and inverse Compton scattering, the
photons survive for a time determined by either the energy loss rate
for compton scattering or the interaction rate for ordinary pair
production in the gas. During this time they can produce light nuclei
by photodisintegration. The electrons and positrons give rise to first
generation photons as a result of inverse Compton scattering. These
first generation photons then produce the second generation photons
and so on. Each generation of photons is shifted to low energies
because the inverse Compton scattering is in the Thompson regime and
only one to two generations of photons are sufficiently energetic to
induce cascade nucleosynthesis.

At $T\la 10^{-4}$ keV, when interaction times become larger than the
Hubble time, only the zero-generation photons are produced and the
effectiveness of the cascade nucleosynthesis diminishes as $T$ drops.

A detailed numerical cascade calculation including all the effects has
been recently performed by Protheroe {\it et al.}, \re{detail} to find
the number of deuterium and $^3$He nuclei produced by the cascade
initiated at a redshift $z$. It was shown that because of the almost
instant formation of the zero-generation spectrum, the exact shape of
the $\gamma$-ray or electron injection spectrum is of no consequence
for further cascade development, and only the total amount of injected
energy is relevant. The epoch of cascade nucleosynthesis is limited by
$T_{{\rm max}}$ and $T_{{\rm min}}$. The maximum temperature is
determined by the condition that the maximum energy of photons in the
cascade spectrum, $E_{{\rm max}}$, must be larger than the threshold
for D or $^3$He production on $^4$He, $E_{{\rm max}}\simeq 20$
MeV. This condition result in $T_{{\rm max}}\simeq 0.57$ keV. The
effectiveness of D and $^3$He production decreases as $T$
decreases. The reason for that is not the decrease of the density of
$^4$He, but rather the decrease in number of low-energy photons in the
cascade. In fact, the lower the temperature, the higher the photon
energies in the cascade and therefore a smaller fraction of the
photons in cascade nucleosynthesis \re{detail}.

At $T\la T_{{\rm min}}\simeq 2.4\times 10^{-4}$ keV the cascade
photons can be directly observed and the upper limit for the isotropic
$\gamma$-ray flux at $10-200$ MeV is more restrictive for the cascade
production than nucleosynthesis. Therefore the most effective epoch
for cascade nucleosynthesis corresponds to temperatures in the range
$10^{-4}-0.5$ keV \re{detail}.

It was also shown that the role of
$\gamma\gamma\rightarrow\gamma\gamma$ scattering is important for
epochs with temperatures $T\ga 1.2\times 10^{-2}$ keV when the
scattered photons do not interact again with the target photons, but
are just redistributed over the spectrum producing a small bump before
the cut-off energy $E_{{\rm max}}$.

\vspace{36pt}
\centerline{\bf IV. APPLICATION AND DISCUSSION}
\vspace{24pt}

{}From the detailed analysis performed in \re{detail},
the number of $^3$He and D nuclei produced by a single cascade of
total energy $E_0$  is ${\cal N} (^3$He$)\sim(0.1-1)\:E_0$ GeV$^{-1}$ and
${\cal N} ({\rm D})\sim (1-6)\times 10^{-3}\:E_0$ GeV$^{-1}$.

Let us imagine that eternal self-annihilations of light photinos into
lepton pairs are the only source of energetic cascades. In such a
case, the total fraction ($^3$He+D)/H at the present epoch reads
\begin{eqnarray}
\label{final}
\frac{^3{\rm He}+{\rm D}}{{\rm H}}&=&f_c\:\left(\frac{
m_{\ph}}{{\rm GeV}}\right)\:\int\:dt\:\frac{\dot{n}^c_{\ph}}{n_{{\rm H}}}
\:\left[{\cal N} (^3{\rm He},t)+{\cal N} ({\rm D},t)\right]\nonumber\\
&=&-f_c\:\left(\frac{
m_{\ph}}{{\rm GeV}}\right)\:\int\:\frac{dt}{n_{{\rm H}}}
\:\left[{\cal N} (^3{\rm He},t)+{\cal N} ({\rm D},t)\right]
\:\langle \sigma_{{\rm SA}} |v|\rangle Y_{\infty}^2\:s^2,
\end{eqnarray}
where $f_c$ is the fraction of mass $m_{\ph}$ transferred to cascade
energy, ${\cal N} (^3{\rm He},t)$ and ${\cal N} ({\rm D},t)$ are the
number of $^3$He and D nuclei produced at the time $t$ per GeV by the
total electromagnetic cascade \re{detail}, $n_{\rm H}$ is the hydrogen
number density per comoving volume and $\dot{n}^c_{\ph}$ represents
the time derivative of the photino number density per comoving volume.

Photino self-annihilations in which the final state is a
lepton-antilepton pair involve the $t$-channel exchange of a virtual
slepton between the photinos, producing the final fermion-antifermion
pair. In the low-energy limit the mass $m_{\tilde{l}}$ of the slepton
is much greater than the total energy exchanged in the process and the
photino-photino-fermion-antifermion operator appears in the low-energy
theory as a coefficient proportional to $e_i^2/m^2_{\tilde{l}}$, with
$e_i$ the charge of the final-state fermion. Also, as there are two
identical fermions in the initial state, the annihilation proceeds as
a $p$-wave, which introduces a factor $v^2$ in the low-energy cross
section \re{gold}. The resultant low-energy photino self-annihilation
cross section is \re{jungman}
\begin{equation}
\label{crossection}
\langle \sigma_{{\rm SA}} |v|\rangle =8\pi\:\alpha_{{\rm em}}^2\:\sum_i\:
e_i^4\:\frac{m_{\ph}^2}{m_{\tilde{l}}^4}\:\frac{v^2}{3}\simeq 2.3\times
 10^{-11}\:\left(\frac{T}{m_{\ph}}\right)\:\left(\frac{m_{\ph}}{{\rm GeV}}
\right)^2\:\left(\frac{m_{\tilde{l}}}{100\:\:{\rm GeV}}\right)^{-4}\:\:
{\rm mb},
\end{equation}
where we have used for the relative velocity $v^2=6/x$ and we have
summed over $e$ and $\mu$ assuming a common slepton mass
$m_{\tilde{l}}$ for selectron and smuon scalar fields.

Recalling that for a radiation-dominated Universe
$t=0.301\:g_*^{-1/2}\left(M_{{\rm Pl}}/T^2\right)$, assuming baryonic
mass is 77\% hydrogen by mass, and making use of \eqr{start}, we have
numerically evaluated the time-integral of the right-hand side of
\eqr{final} and obtained the present total fraction ($^3$He+D)/H
generated by the electromagnetic showers induced by eternal photino
self-annihilations into lepton-antilepton pairs
\begin{equation}
\label{con}
\frac{^3{\rm He}+{\rm D}}{{\rm H}}\simeq 9.4\times 10^{-12}\:
\left(\frac{f_c}{0.77\:\Omega_{{\rm B}}}\right)\:
\left(\Omega_{\ph}\:h\right)^2\:
\left(\frac{m_{\tilde{l}}}{100\:\:{\rm GeV}}\right)^{-4}.
\end{equation}
The upper limit inferred from measurements of $^3$He in meteorites and
the solar wind making assumptions about stellar processing and
galactic chemical evolution is ($^3$He+D)/H$\la 1.1\times 10^{-4}$
\re{limit}. This bound translates from \eqr{con} into a lower limit on
the slepton mass\footnote{Note, however, that very recent measurements
of D/H are closer to $(2-5)\times 10^{-4}$ \re{new} and, if such
higher values for ($^3$He+D)/H were adopted, the upper limits we
derive would be correspondingly higher.}
\begin{equation}
m_{\tilde{l}}\ga 1.7\:\left(\frac{f_c}{0.77\:\Omega_{{\rm B}}}\right)^{1/4}\:
\left(\Omega_{\ph}\:h \right)^{1/2}\:\:{\rm GeV}.
\end{equation}
We notice that this lower bound on the slepton mass is {\it
independent} of the photino mass and thus holds for the entire range
of cosmological interest of the photino mass $1.6\la r\la 2$
\re{kolb}. Since the lower limit on $m_{\tilde{l}}$ given in Eq. (11)
is weaker than any other present experimental lower bound on slepton
masses, we may conclude that eternal self-annihilations of light
photinos do not jeopardize the successful predictions of primordial
nucleosynthesis.

Let us now generalize our analysis to the case in which the relic
abundance of a generic light dark matter candidate $X$ is determined
by some processes other than its self-annihilation cross-section into
leptons, which we parameterize as $\langle \sigma_{{\rm SA}} |v|\rangle
=\sigma_0\:(T/m_X)^n$, where we have assumed that either $s$-wave
($n=0$) or $p$-wave ($n=1$) annihilations dominate\footnote{This may happen, 
for instance,  when explicit sfermion mixing \re{mix} or $CP$-violating phases \re{cp} 
in low-energy supersymmetric models greatly enhance the self-annihilation 
cross-section of the lightest
supersymmetric particles into fermions 
removing the $p$-wave suppression.}; it is
straightforward to generalize to the case where both contribute. For
instance, one can envisage the situation in which the dark matter
component $X$ is slightly degenerate with a particle $X^\prime$ and
that interconversion processes $X\longleftrightarrow X^\prime$ keep $X$ in
equilibrium even after self-annihilations have frozen out, regardless
of the strength of the interactions governing the $X^\prime$
abundance. This case is rather different from the one analyzed in
\re{residual} where it was assumed that the same cross section
determines both the self-annihilation rate and the relic abundance of
the $X$-particles.

By making use of Eqs.\ (\ref{bol}) and (\ref{final}), we have
numerically evaluated the present fraction of ($^3$He+D)/H generated
by electromagnetic cascades induced by eternal self-annihilations of
$X$'s into leptons. Adopting the upper limit ($^3$He+D)/H$\la
1.1\times 10^{-4}$ \re{limit} we find
\begin{eqnarray}
\left(\frac{\sigma_0}{10^{-38}\:{\rm cm}^{-2}}\right)\:
\left(\frac{m_{X}}{{\rm GeV}}\right)^{-1}&\la& 2.7\times 10^7\:
\left(\frac{f_c}{0.77\:\Omega_{{\rm
B}}}\right)^{-1}\:\left(\Omega_{X}\:h\right)^{-2}\:\:\:\:\:{\rm
for}\:\: n=0,\nonumber\\ \left(\frac{\sigma_0}{10^{-38}\:{\rm
cm}^{-2}}\right)\: \left(\frac{m_{X}}{{\rm GeV}}\right)^{-2}&\la&
6.6\times 10^{12}\: \left(\frac{f_c}{0.77\:\Omega_{{\rm
B}}}\right)^{-1}\:\left(\Omega_{X}\:h\right)^{-2}\:\:\:\:{\rm for}\:\:
n=1.\nonumber\\ &&
\end{eqnarray}
Once the model-dependent cross section $\sigma_0$ is known, one can
apply the above limits to find the appropriate bounds on the parameter
space of the model.

So far we have been assuming that the generic dark matter component
$X$ eternally self-annihilates only into lepton-antilepton pairs. When
hadronic channels are open, D is produced by hadronic showers and this
requires reconsideration of the constraint derived above. In fact,
even if light photinos self-annihilate only into lepton pairs, the
resulting electromagnetic cascades will be effectively hadronic if
$E_\gamma\varepsilon_\gamma\ga$ GeV$^2$, where $E_\gamma$ is the
typical energy of a cascade photon and $\varepsilon_\gamma$ is the low
energy of a blackbody photon; furthermore there is always a $\sim$ 1\%
probability for the virtual decay photon to convert into a
quark-antiquark pair over threshold. Hence hadronic showers will be
generated if $m_X\ga 1$ GeV even if $X$-particles do not have specific
self-annihilation channels into quark pairs (see Reno and Seckel
\re{residual}).

As discussed in detail by Dimopoulos {\it et al.}\ \re{dim}, the main
effect of hadronic cascades is the destruction of the ambient $^4$He
and the creation of D (while it is photodestroyed by electromagnetic
showers), $^3$He, $^6$Li, and $^7$Li. Even though the final answer on
the light element abundances may be provided only by an extensive
numerical analysis involving both hadronic and electromagnetic
cascades, since the constraints on the primordial abundance of $^6$Li
and $^7$Li are so stringent \re{sarkar} one might expect that energy
released by self-annihilations of $X$-particles under the form of
quark-antiquark pairs must be very small in order not to overproduce
lithium. This suppression may happen either imposing that the
$X$-particle is sufficiently light so that emission of quark pairs is
phase-space suppressed or assuming that the self-annihilation cross
section is enough suppressed, which may happen if the virtual
particles mediating the process are very heavy.

In conclusion, we have investigated the effect on big-bang
nucleosynthesis predictions of a very light photino predicted in
low-energy supersymmetry models in which dimension-3 supersymmetric
breaking operators are suppressed \re{glennys}. We have found that
eternal photino self-annihilations into fermion pairs that take place
around $10^5$ to $10^6$ sec after the big-bang cannot significantly
alter the primordial abundances of light elements when taking into
account the present experimental limits on the slepton masses.

We have also generalized our study to the case of a generic dark
matter component $X$ whose relic abundance is not determined by the
self-annihilation cross section into leptons, but by some other
processes, {\it e.g.,} co-annihilations with a slightly heavier
particle $X^\prime$. Limits on the self-annihilation cross section and
mass of the $X$-particle have been given by requiring that
electromagnetic cascades generated by the eternal self-annihilations
of the $X$-particle do not jeopardize the predictions of the big-bang
nucleosynthesis.

\vspace{36pt}

\centerline{\bf ACKNOWLEDGMENTS}

EWK and AR are supported by the DOE and NASA under Grant NAG5--2788.

\frenchspacing
\def\prpts#1#2#3{Phys. Reports {\bf #1}, #2 (#3)}
\def\prl#1#2#3{Phys. Rev. Lett. {\bf #1}, #2 (#3)}
\def\prd#1#2#3{Phys. Rev. D {\bf #1}, #2 (#3)}
\def\plb#1#2#3{Phys. Lett. {\bf #1B}, #2 (#3)}
\def\npb#1#2#3{Nucl. Phys. {\bf B#1}, #2 (#3)}
\def\apj#1#2#3{Astrophys. J. {\bf #1}, #2 (#3)}
\def\apjl#1#2#3{Astrophys. J. Lett. {\bf #1}, #2 (#3)}
\begin{picture}(400,50)(0,0)
\put (50,0){\line(350,0){300}}
\end{picture}

\vspace{0.25in}

\def\labelenumi{[\theenumi]}

\begin{enumerate}

\item\label{haber} H. E. Haber and G. L. Kane, Phys. Rep. {\bf 117}, 75 (1985) .

\item\label{glennys} G. R. Farrar and A. Masiero, Rutgers Univ.
Technical Report RU-94-38 (hep-ph/9410401); G. R. Farrar,
Phys. Rev. {\bf D51}, 3904 (1995); Rutgers Univ.  Technical Report
RU-95-17 (hep-ph/9504295); Rutgers Univ. Technical Report RU-95-25
(hep-ph/9508291); Rutgers Univ. Technical Report RU-95-26; Rutgers
Univ. Technical Report RU-95-73; Rutgers Univ. Technical Report
RU-95-74.

\item\label{kolb} G. R. Farrar and E. W. Kolb, Phys. Rev. D (in press). 

\item\label{lindley} D. Lindley, Mon. Not. R. Astron. Soc. {\bf 188},
15 (1979); Astrophys. J. {\bf 294}, 1 (1985); J. Audoze, D. Lindley
and J. Silk, {\it ibid.} {\bf 293}, L53 (1985).

\item\label{residual} J. S. Hagelin and R. J. D. Parker,
Phys. Lett. {\bf B215}, 397 (1988) and Nucl. Phys. {\bf B329}, 464
(1990); M. H. Reno and D. Seckel, Phys. Rev. {\bf D37} 3441, (1988) ;
J. A. Frieman, E. W. Kolb and M. S. Turner, Phys. Rev. {\bf D41}, 3080
(1990).

\item\label{griest} K. Griest and D. Seckel, Phys. Rev. {\bf D43},
3191 (1991).

\item\label{kt} See, {\it e.g.}, E. W. Kolb and M. S. Turner, {\it The
Early Universe} (Addison-Wesley), Redwood City, CA, 1989), Chapter 5.
 
\item\label{preparation} D. J . Chung, G. R. Farrar and E. W.  Kolb,
in preparation.

\item\label{sarkar} For a nice review on big-bang nucleosynthesis and
relative constraints coming from production of electromagnetic
cascades, see S. Sarkar, OUTP-95-16P preprint, submitted to Rep. on
Prog. in Phys., and referenced therein.

\item\label{elastic} A. Zdziarski and R. Svensson, Astrophys. J. {\bf
344}, 551 (1989).

\item\label{ellis} J. Ellis {\it et al.}, Nucl. Phys. {\bf 373}, 399 (1992).

\item\label{detail} R. J. Protheroe, T. Stanev and V. S. Berezinsky,
Phys. Rev. {\bf D51}, 4134 (1995).

\item\label{gold} H. Goldberg, Phys. Rev. Lett. {\bf 50},  1419  (1983).

\item\label{jungman} See, for instance, G. Jungman, M. Kamionkowski
and K. Griest, SU-4240-605 preprint, submitted to Phys. Rep.

\item\label{limit} C. J. Copi, D. N. Schramm and M. S. Turner, Science
{\bf 267}, 192 (1995).

\item\label{new} R. F. Carswell {\it et al.},
Mon. Not. R. Astron. Soc. {\bf 268}, L1 (1994); A. Songalia {\it et
al.}, Nature (London) {\bf 368}, 599 (1994).

\item\label{dim} S. Dimopoulos {\it et al.}, Nucl. Phys. {\bf 311},
699 (1989).

\item\label{mix} T. Falk, R. Madden, K.A. Olive and M. Srednicki, 
Phys. Lett. {\bf B318},  354 (1993).

\item\label{cp} T. Falk,  K.A. Olive and M. Srednicki, Phys. Lett. {\bf B354},
 99  (1995).

\end{enumerate}

\end{document}